\documentclass[10pt]{iopart}
\usepackage{iopams}
\usepackage{graphicx}
\usepackage{float}

\begin{document}

\title[Windproofing LIGO]{Towards windproofing LIGO: Reducing the effect of wind-driven floor tilt by using rotation sensors in active seismic isolation.}

\author{Michael P. Ross$^{1*}$, Krishna Venkateswara$^{1}$,  Conor Mow-Lowry$^2$, Sam Cooper$^2$, Jim Warner$^3$, Brian Lantz$^4$, Jeffrey Kissel$^3$ , Hugh Radkins$^3$, Thomas Shaffer$^3$, Richard Mittleman$^5$,  Arnaud Pele$^6$, and Jens Gundlach$^1$}
\ead{mpross2@uw.edu$^*$}
\address{$^1$Center for Experimental Nuclear Physics and Astrophysics, University of Washington, Seattle, Washington 98195, USA, 
\\$^2$University of Birmingham, Birmingham B15 2TT, United Kingdom, 
\\$^3$LIGO Hanford Observatory, Richland, Washington 99352, USA, 
\\$^4$Stanford University, Stanford, California 94305, USA, 
\\$^5$Massachusetts Institute of Technology, Cambridge, Massachusetts 02139, USA, 
\\$^6$LIGO Livingston Observatory, Livingston, Louisiana 70754, USA}
\maketitle


\begin{abstract}
Modern gravitational-wave observatories require robust low-frequency active seismic isolation in order to keep the interferometer at its ideal operating conditions. Seismometers are used to measure both the motion of the ground and isolated platform. These devices are susceptible to contamination from ground tilt at frequencies below 0.1 Hz, particularly arising from wind-pressure acting on building walls. Consequently, during LIGO's first observing run both observatories suffered significant downtime when wind-speeds were above 7 m/s. We describe the use of ground rotation sensors at the LIGO Hanford Observatory to correct nearby ground seismometers to produce tilt-free ground translation signals. The use of these signals for sensor correction control improved low-frequency seismic isolation and allowed the observatory to operate under wind speeds as high as $15-20$ m/s.
\end{abstract}

\submitto{\CQG}

\section{Introduction}

In recent years, gravitational waves have become a novel method to observe the universe. During the first three observation runs of the Laser Interferometer Gravitational-wave Observatory (LIGO) \cite{aligocqg}, gravitational-waves (GW) have been observed from multiple compact binary mergers \cite{GWTC, GW190425, GW190412}. These observations have furthered our understanding of a variety of phenomena from the equation of state of neutron stars \cite{NSEoS} and the creation of heavy elements \cite{heavyElements} to populations of black holes \cite{GWTC} and the nature of gravity \cite{speedGW}. 

In order to detect faint GW signals, The LIGO interferometers must be decoupled from all environmental effects. The most immediate of these is ambient seismic motion. In order to isolate from ground motion, LIGO deploys a multi-stage seismic isolation system \cite{ISIarticle}. Although seismic motion can couple into the interferometers gravitationally \cite{terrestrial}, here we will focus only on direct mechanical coupling between the ground and the interferometer's optics.

The duty-cycles of the LIGO Hanford Observatory (LHO) and the LIGO Livingston Observatory (LLO) during the first observing run were 64.4\% and 57.3\%, respectively, with significant downtime due to high wind (3.9\%, 6.5\%), microseismic motion (2.5\%, 5.8\%), and earthquakes (2.6\%, 6.9\%) \cite{LIGODetectorsO1}. As GW events are transient in nature, any increase in observing time increases the number of detections. Even when the interferometers remain operational, excess noise has been observed during elevated seismic and atmospheric activity. \cite{detChar}

\begin{figure}[!h]
\begin{center}
\includegraphics[width=0.65\textwidth]{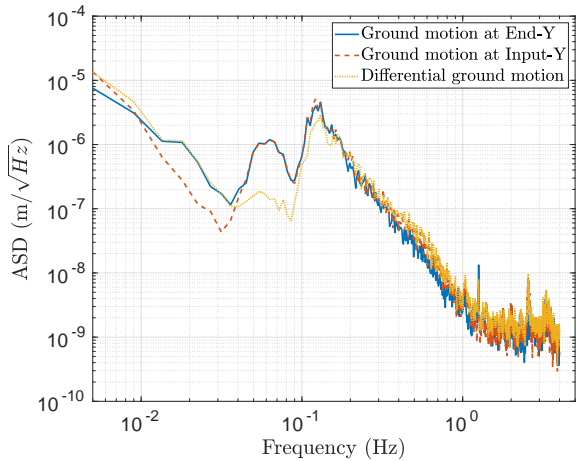}
\end{center}
\caption{Amplitude spectral density (ASD) of the ground motion along the Y-arm at LHO, measured near the Input-Y and End-Y test masses and the differential motion between the two. The primary and secondary microseismic peaks are clearly visible between $0.05$ to $0.5$~Hz.\label{GroundMotion}}
\end{figure}

To keep a LIGO interferometer ``locked'' (operating at its optimum point) the net differential motion of the suspended test-masses in the two arm-cavities must be less than $10^{-11}$ m RMS. The differential ground motion over the observatory's $4$-km long arms has a strong frequency dependence and varies with time. Fig.~\ref{GroundMotion} shows an example of the horizontal ground motion measured at LHO  when the wind speeds were below $3$ m/s. As will be seen later, wind-induced ground tilt can dominate the observed
ground motion below 0.1 Hz \cite{venk2016}. These tilts arise from deformations of the observatory's floor due to wind-driven pressure fluctuations on the walls of the building. Between $0.05$ to $0.5$~Hz, the ground motion is largely dictated by Rayleigh waves arising from ocean waves interacting with the sea floor \cite{Longuet-Higgins1}. The $10-15$ second period ocean waves produce the so-called primary, secondary, and tertiary microseism peaks through interference effects. The amplitude of the secondary is usually the largest and can vary from $\sim$ $0.1-3$~$\mu$m depending on oceanic activity and geographic location. 

The differential ground motion over the 4-km arm is reduced significantly between $0.05 - 0.2$~Hz, due to the fact that the wavelength of the seismic waves at these frequencies is $\sim$ $40$~km. Additionally, at the diurnal and semi-diurnal frequency, the earth tides  can produce between $100-250$~$\mu$m of differential displacement over the arms. To operate the interferometer in such an environment, multiple feedback loops are deployed to minimize both the translational and angular motion of the test masses.

The data and analysis in this paper is limited to LHO, which was the only site that received tilt-sensors between the first (O1) and second (O2) observing runs. The ground motion at the two LIGO sites are notably different (due to the geographical location and geological properties) which makes it difficult to immediately apply the following results to LLO. Nevertheless, similar schemes are deployed at LLO and could be applied to other terrestrial gravitational-wave observatories.

\subsection{Platform Control configuration}
The LIGO seismic isolation systems consist of multiple suspended platforms and use seismometers to measure motion with reference to an inertial frame, see Ref.~\cite{ISIarticle} for details. In general, the suspended platforms can be controlled by feedback using on-board sensors or by feedforward using sensors on the ground. When viewed locally, these two approaches are nearly equivalent. However, using on-board sensors for feedback is slightly superior as the motion sensed at the ground and the motion affecting the platform may be incoherent. At many frequencies, specifically at the microseism frequencies (0.05-0.5~Hz), the ambient seismic motion is sourced by many disparate emitters. This causes the observed motion to be location dependent. Thus, the coherence between measurements at two locations decreases if their separation is greater than a seismic wave's wavelength. This wavelength is dependent on both the seismic wave's frequency and the local geology.~\cite{seismologyBook}

In the first observing run, only on-board inertial sensors were used for feedback down to approximately 50 mHz. This scheme is referred to as the O1 configuration. Below 50 mHz, relative position sensors, which sense motion relative to two isolation stages, where used. Between O1 and O2, two beam-rotation-sensors (BRS) \cite{brs} were installed at the end-stations and used to subtract the tilt-coupling of the ground seismometers \cite{venk2016}. The tilt sensed by the corner-station seismometer was significantly less than the end-station seismometers under similar wind-speeds. This was due to the greater distance from corner station seismometer to the corner station walls than was achievable at the end stations. The tilt-subtracted seismometers at the end-stations and the tilt-free seismometer at the corner station form a set of low-tilt seismometers. For the second observing run, the low-tilt ground seismometers were used for sensor correction (detailed in Section \ref{model}) at intermediate frequencies and on-board sensors were used for feedback at high frequencies. At low frequencies, the relative position sensors were used to lock to the ground. This is referred to as the O2 configuration. 

\begin{figure}[!h]
\begin{center}
\includegraphics[width=0.75\textwidth]{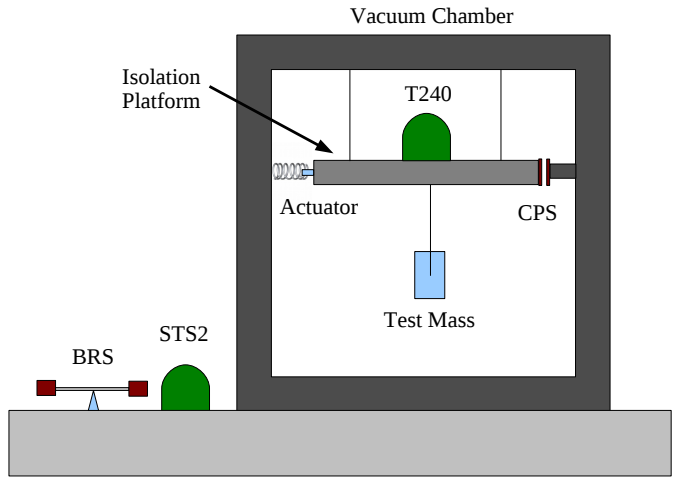}
\end{center}
\caption{Schematic representation of the Internal Seismic Isolation platform showing the on-board seismometer (T240), capacitative position sensor (CPS), the ground seismometer (STS2), and the ground tilt-sensor (BRS). \label{Schematic}} 
\end{figure}

The drawback of the O1 configuration is that it increased platform motion dramatically under windy conditions due to increased ground tilt. It also failed to take advantage of the correlated ground motion over the $4$-km arms in the 50-150 mHz band and high correlation across a wide frequency range over the tens of meters that separate the corner station chambers.

The goal of the O2 configuration is to decrease the residual platform motion below 0.1 Hz during high winds. This decreased motion decreases the demand on the downstream control loops \cite{ASC} which keep the interferometer locked. With the O1 configuration, these control loops were often overwhelmed during high winds which would cause the interferometer to loose lock.

\section{Simple Analytical Model}\label{model}

The active isolation for the LIGO test-masses employs two main subsystems- the Hydraulic External Pre-Isolation (HEPI) and the Internal Seismic Isolation (ISI). For a detailed description see \cite{ISIarticle}. The ISI is a dual-stage active isolation system from which the test mass assembly is suspended. Here we will only discuss the first stage of the ISI as it accounts for the majority of the low-frequency isolation performance and its controls were changed substantially between O1 and O2. Fig.~\ref{Schematic} shows a simplified schematic of the ISI which includes the on-board seismometer (T240), capacitative position sensor (CPS), the ground seismometer (STS2), and the ground tilt-sensor (BRS).

Following the procedure outlined in Ref. \cite{lantz2008}, we developed a simplified, two degree of freedom, analytical model of the ISI's first stage. While this model lacks the ability to incorporate cross-couplings (unlike a numerical model), it is instructive in understanding the main features of the system and in guiding the development of the control filters. Additionally, cross-couplings do not significantly influence the tilt-sensor implementation scheme.

\begin{figure}[!h]
\begin{center}
\includegraphics[width=0.5\textwidth]{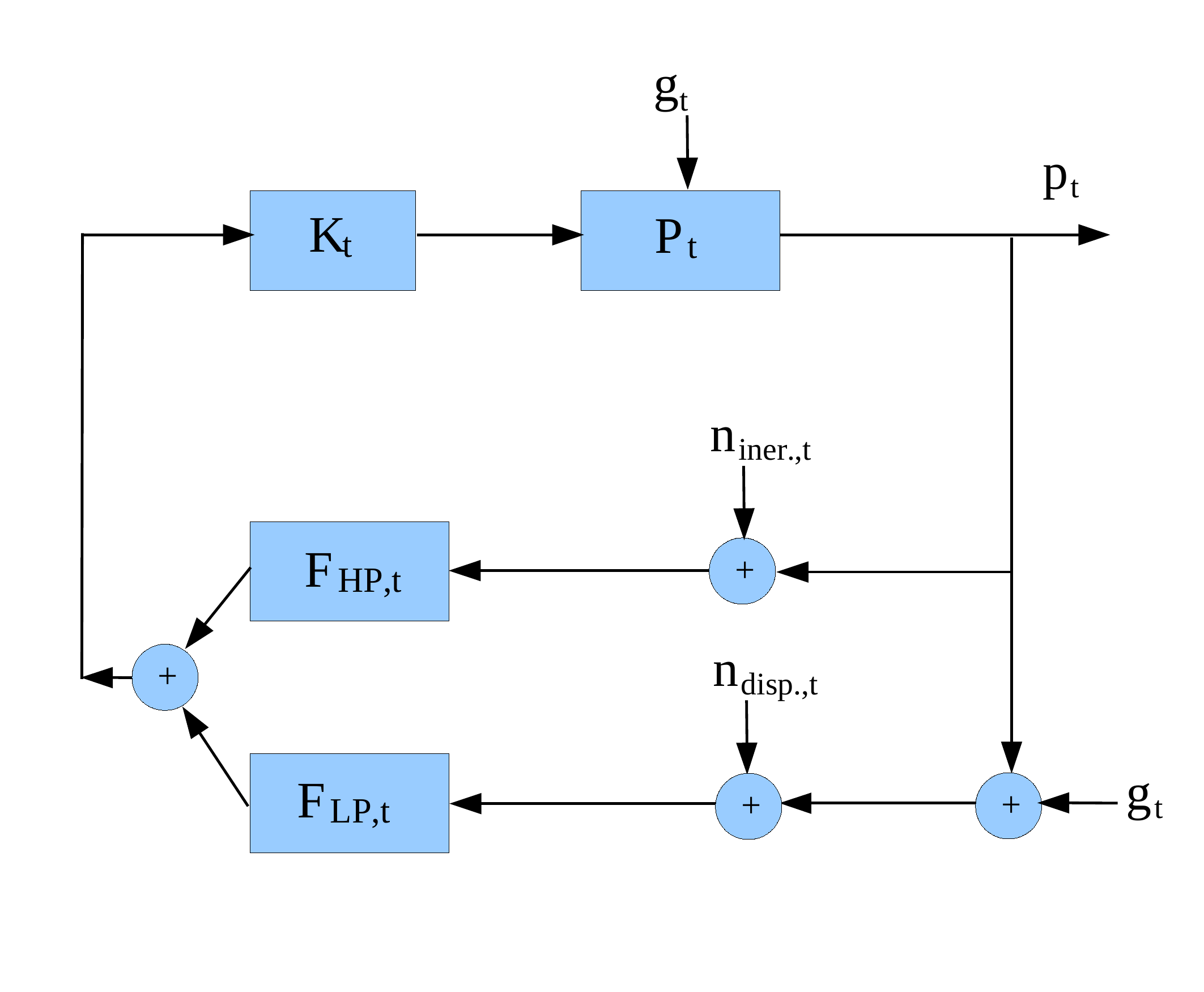}
\end{center}
\caption{Model of the tilt control loop for the first stage of the ISI, where $g_{t}$ is the ground tilt, $p_t$ is the platform tilt, $n_{iner.,t}$ and $n_{disp.,t}$ are the tilt sensor noises for the on-board T240 pair and position sensor respectively, $F_{HP,t}$ and $F_{LP,t}$ are respectively high pass and low pass filters, $K_t$ is the feedback filter, and $P_t$ is the platform tilt transfer function. \label{TiltLoop}}
\end{figure}

\begin{figure}[!h]
\begin{center}
\includegraphics[width=0.5\textwidth]{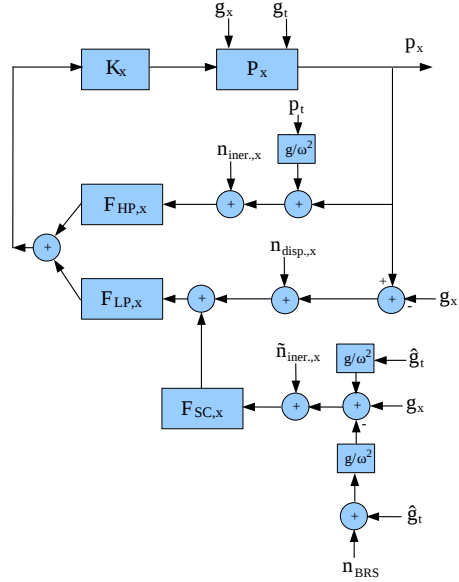}
\end{center}
\caption{Model of the translation control loop for the first stage of the ISI, where $g_{t}$, $\hat{g}_{t}$, and $g_{x}$ are the ground tilt at the platform, ground tilt at the ground seismometer, and ground translation respectively, $p_x$ and $p_t$ are the platform translation and tilt, $n_{iner.,x}$, $\tilde{n}_{iner.,x}$, $n_{disp.,x}$ and $n_{BRS}$ are the sensor noises for the on-board T240, ground STS2, position sensor, and BRS respectively, $F_{HP,x}$, $ F_{LP,x}$, and $ F_{SC,x}$ are respectively a high pass filter, a low pass filter, and the sensor correction filter, $K_x$ is the feedback filter, and $P_x$ is the platform translation transfer function. \label{TransLoop}}
\end{figure}

Fig.~\ref{TiltLoop} and Fig.~\ref{TransLoop} show control loop diagrams for the tilt and translation degrees of freedom, respectively. These show that the control is comprised of the inertial platform sensors at high frequency and the relative position sensors at low frequency. For the translational degree of freedom there is an additional branch which displays sensor correction. Sensor correction is the process of adding a low-tilt ground inertial sensor to remove the ground contribution of the position sensor. This yields a measurement of the inertial motion of the platform at lower frequencies than is obtainable with the on-board inertial sensors.

In the limit of infinite loop gains and at frequencies where the mechanical dynamics of the platform are unimportant, the residual platform motion is described by relatively simple equations.

The platform tilt arising from ground tilt is
\begin{equation}
p_t= F_{LP,t} \cdot g_t.
\label{eq4}
\end{equation}
and the platform tilt due to the sum of sensor noises is
\begin{equation}
p_t= -F_{HP,t} \cdot n_{iner.,t} -  F_{LP,t} \cdot n_{disp.,t}.
\label{eq5}
\end{equation}
where $g_{t}$ is the ground tilt, $p_t$ is the platform tilt, $ F_{LP,t}$ and $ F_{HP,t}$ are, respectively, a low pass and high pass filters, and $n_{iner.,t}$ and $n_{disp.,t}$ are the tilt  sensor noise for the on-board seismometer pair and position sensor, respectively.

Similarly, the platform translations sourced by ground translations can be expressed as
\begin{equation}\label{eq1}
p_x=F_{LP,x} \cdot (1-F_{SC,x}) \cdot g_x.
\end{equation}
where $g_{x}$ is the ground translation, $p_x$ is the platform translation,  $ F_{LP,x}$ and $ F_{SC,x}$ are respectively a low pass filter and the sensor correction filter.

Ground tilt can produce platform translation through the following two terms:
\begin{equation}\label{eq2}
p_x= - F_{LP,x} \cdot F_{SC,x} \cdot \frac{g}{\omega^2} \cdot \hat{g}_t - F_{HP,x} \cdot \frac{g}{\omega^2} \cdot F_{LP,t} \cdot g_t.
\end{equation}
where $g_t$ is the ground tilt at the platform, $\hat{g}_t $ is the ground tilt at the ground seismometer, $p_x$ is the platform translation, $F_{HP,x}$, $ F_{LP,x}$, and $ F_{SC,x}$ are respectively a high pass filter, a low pass filter, and the sensor correction filter, $g$ is the gravitational acceleration, and $\omega$ is the frequency of motion. The first term is due to the tilt-sensitivity of the ground seismometer while the second is due to the platform tilt being sensed as translation by the platform seismometer. 

When a tilt sensor is used to subtract the tilt contamination from the ground sensor, the ground tilt in the first term is replaced with the level of noise in the tilt-subtracted channel. Ideally this would be dominated by the tilt sensor and the seismometer sensor noise. In reality, the tilt-subtracted residual signal is limited by differences in tilt experienced by the two sensors. Hence a signal-dependent 'noise' is added to the tilt-subtracted channel. Eq.~\ref{eq2} is then modified to be: 
\begin{equation}
p_x= - F_{LP,x} \cdot F_{SC,x} \cdot \frac{g}{\omega^2} \cdot n_{TS} - F_{HP,x} \cdot \frac{g}{\omega^2} \cdot F_{LP,t} \cdot g_t.
\label{eq2a}
\end{equation}
where the noise in the tilt-subtracted channel $n_{TS}$ is modelled as
\begin{equation}
n_{TS}= -n_{BRS} +\alpha \cdot \hat{g}_t
\label{eq2b}
\end{equation}
and $\alpha$ is the residual tilt factor which in the current sensor arrangement is  $\sim 0.1$. Decreases in the residual tilt would allow the improvements detailed in the following to be extended to higher wind speeds ($>15$ m/s).

The sum of sensor noises affecting the platform translation can be written as
\begin{eqnarray}\label{eq3}
p_x= & -F_{HP,x} \cdot n_{iner.,x} - F_{LP,x} \cdot n_{disp.,x} \nonumber \\
& - F_{LP,x} \cdot F_{SC,x} \cdot \tilde{n}_{iner.,x} + F_{HP,x}\cdot \frac{g}{\omega^2} \cdot F_{HP,t} \cdot n_{iner.,t}\\
& +  F_{HP,x} \cdot \frac{g}{\omega^2} \cdot F_{LP,t} \cdot n_{disp.,t} \nonumber
\end{eqnarray}
where $p_x$ is the platform translation, $n_{iner.,x}$, $\tilde{n}_{iner.,x}$, and $n_{disp.,x}$ are respectively the translation noise for the on-board T240, ground STS2, and CPS, $n_{iner.,t}$ and $n_{disp.,t}$ are the tilt noise for the on-board T240 pair and CPS respectively. Note that the first three terms are due to the translational sensor performance while the last two are due to tilt contamination of the on-board seismometers. These last two terms couple the performance of the tilt degree of freedom to the translational performance.

\begin{figure}[!h]
\begin{center}
\includegraphics[width=0.65\textwidth]{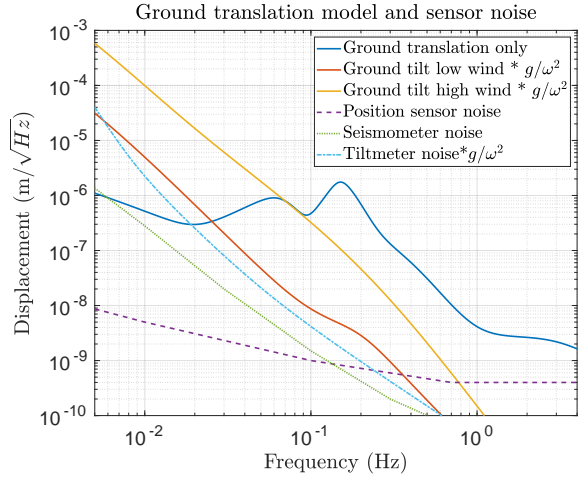}
\end{center}
\caption{ASD of the modeled ground translation, along with low and high tilt models as sensed by a seismometer. Also shown are translational sensor noise models for the position sensor (CPS), seismometer (T240), and tiltmeter (BRS). \label{TransModel}}
\end{figure}

To study the effects of different filters configurations, we construct an idealized model of both the translation and rotational inputs to the system. Fig.~\ref{TransModel} shows the Amplitude Spectral Density (ASD) of the modeled ground translation and ground tilt sensed as translation during low ($0-5$ m/s) and high ($10-12$ m/s) wind conditions along with various sensor noise models. The sensor noise models represent the design sensitivity of each instrument. We assume that the ground translation does not change under different wind conditions, but the increase in the sensed translation is only due to the increased tilt. In reality, the ground motion above $\sim$0.1 Hz increases slightly with wind speed. Additionally, we assume that the platform and the ground seismometer experience the same amplitude, but incoherent, ground tilts as they are located similar distances from the walls of the building. 

\begin{figure}[!h]
\begin{center}
\includegraphics[width=0.65\textwidth]{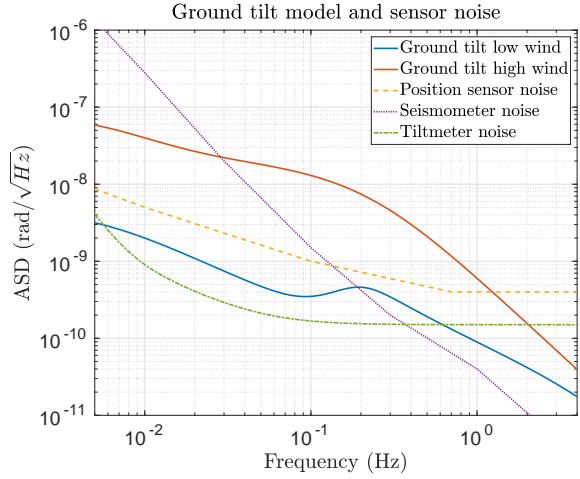}
\end{center}
\caption{ASD of modeled low and high ground tilt along with tilt sensor noise models for the position sensor (CPS), seismometer (T240), and tiltmeter (BRS).\label{TiltModel}}
\end{figure}

Fig.~\ref{TiltModel} shows the ground tilt under low wind and high wind conditions. The ground tilt model was constructed by fitting the tilt spectrum observed by the BRSs at LHO during a given wind speed to an empirically determined model. This was limited to frequencies between 10 mHz $-$ 0.5 Hz to avoid frequencies where the BRS was dominated by instrumental noise. The model was then extrapolated to frequencies above and below this.


\subsection{O1 Low Wind}
\begin{figure}[!h]
\begin{center}
\includegraphics[width=0.65\textwidth]{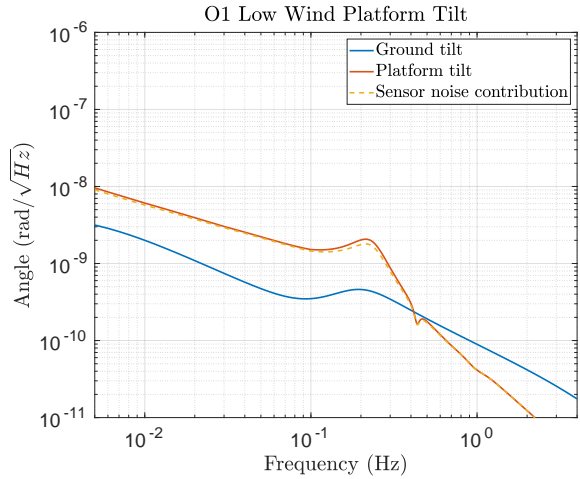}
\end{center}
\caption{ASD of modeled residual platform tilt during low winds. Below $\sim$ $500$ mHz the platform tilt is greater than the ground due to the position sensor noise.\label{PTiltLow}}
\end{figure}

Using these models and Eqs.~\ref{eq1}-\ref{eq5}, we can compute the residual platform motion for the O1 and O2 configurations during low and high winds. Fig.~\ref{PTiltLow} shows the residual platform tilt under low wind conditions with the O1 configuration. The platform tilt is greater than the ground tilt below $\sim$ $500$ mHz due to the position sensor noise.

\begin{figure}[!h]
\begin{center}
\includegraphics[width=0.65\textwidth]{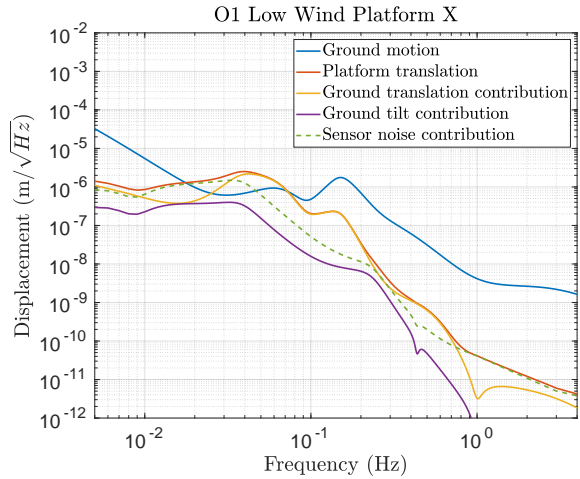}
\end{center}
\caption{ASD of modeled platform translation during low wind in the O1 configuration. The yellow and purple curves represent, respectively, platform translation caused by ground translations and ground tilts while the red curve represent the residual platform translation. \label{PTransLow}}
\end{figure}

Similarly, Fig.~\ref{PTransLow} shows the translation of the platform under low wind conditions with the O1 configuration. At frequencies above $\sim$ $0.65$ Hz, the platform achieves a nearly sensor-noise-limited performance. Below this, the translation of the platform is dominated by residual ground motion leaking into the stop band of the low-pass filter. However, as the sensor noise is only a factor of few below the platform motion, the differential motion between platforms located near one another (such as those in the corner station) is sensor noise limited.

\subsection{O1 High Wind}
\begin{figure}[!h]
\begin{center}
\includegraphics[width=0.65\textwidth]{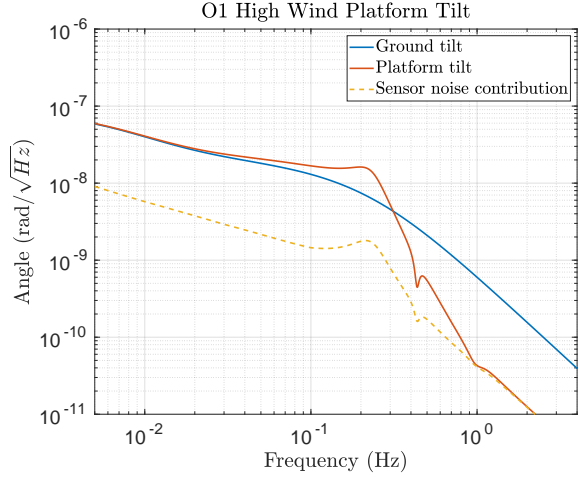}
\end{center}
\caption{ASD of modeled platform and ground tilt during high wind.\label{PTiltWindy}}
\end{figure}

When wind speeds increase, the ground tilt increases significantly as shown in Fig.~\ref{PTiltWindy}. Since the platform is locked to the ground below 30 mHz with the position sensors, this increased ground tilt directly increases the residual platform tilt. Above 1 Hz, the platform achieves similar tilt performance as it does during low winds.

\begin{figure}[!h]
\begin{center}
\includegraphics[width=0.65\textwidth]{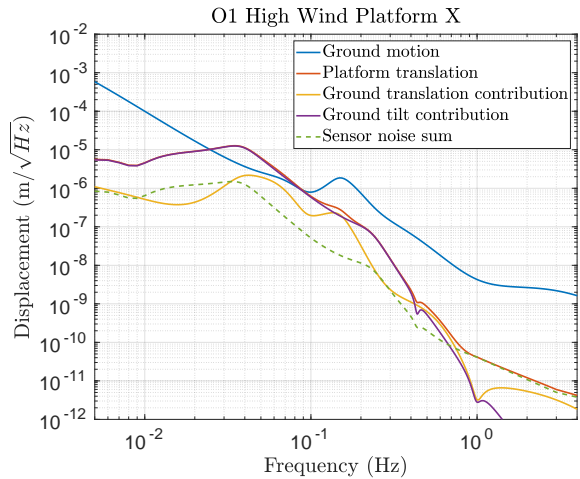}
\end{center}
\caption{Platform translation during high winds in the O1 configuration. The yellow and purple curves represent respectively platform translation caused by ground translations and ground tilts while the red curve represent the residual platform translation.
\label{PTransWindy}}
\end{figure}

As a result of this increased platform tilt, the platform motion increases as shown in Fig.~\ref{PTransWindy}. The tilt contribution is the limiting term at most frequencies below 1 Hz and increases the residual platform motion by a factor of $\sim20$ at $50$ mHz. The increased low-frequency platform translation leads to an increase in the angular motion of the suspended mirrors due to the length to angle cross-coupling of the suspension. This increased angular motion often overloads the downstream control loops \cite{ASC} which fight the test mass angular motion leading to lockloss.

\subsection{O2 Low Wind}
\begin{figure}[!h]
\begin{center}
\includegraphics[width=0.65\textwidth]{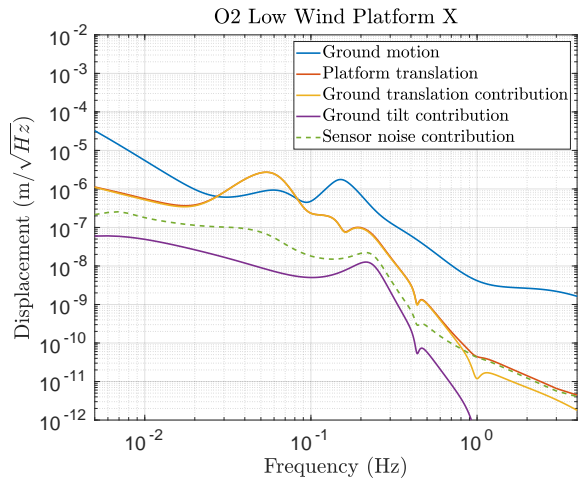}
\end{center}
\caption{Platform translation during low wind in the O2 configuration.The yellow and purple curves represent respectively platform translation caused by ground translations and ground tilts while the red curve represent the residual platform translation.\label{PTransQuietO2}}
\end{figure}

In comparison, Fig.~\ref{PTransQuietO2} shows the platform's translation response under the O2 configuration during low wind. Note that the O2 platform tilt performance is identical to the O1 configuration since the tilt control scheme was not changed. The platform's isolation near the microseism ($40-200$ mHz) and above $1$ Hz is similar to the O1 configuration, but the isolation is degraded between $0.2$ to $0.7$ Hz. The gain peaking near 60 mHz is also comparable to before. However, the sensor noise contribution, mainly comprised of the position sensor tilt, is suppressed significantly at low frequencies. With this configuration the platform motion is dominated by ground motion in this band, which is mostly common over the 4-km long arms as seen in Fig.~\ref{GroundMotion}. Thus, the differential platform motion is reduced as compared to the O1 configuration.

\subsection{O2 High Wind}
\begin{figure}[!h]
\begin{center}
\includegraphics[width=0.65\textwidth]{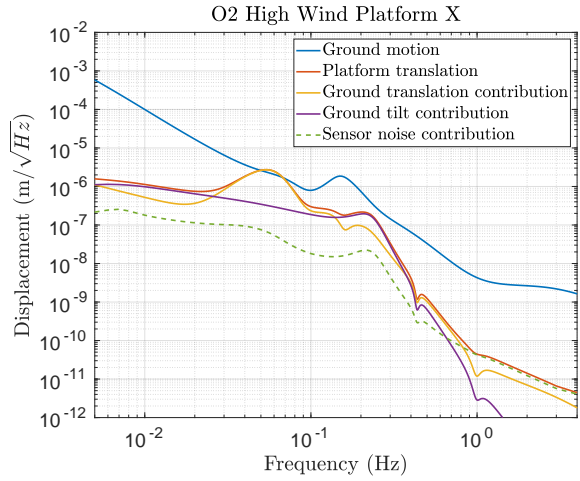}
\end{center}
\caption{Platform translation during high wind in the O2 configuration. The yellow and purple curves represent respectively platform translation caused by ground translations and ground tilts while the red curve represent the residual platform translation.\label{PTransWindyO2}}
\end{figure}

Fig.~\ref{PTransWindyO2} shows the platform performance under windy conditions, where the benefit from the tilt sensor is most visible. The tilt contribution no longer dominates between $30-100$ mHz and is only a factor of $\sim$3 above the ground motion contribution at $8-25$ mHz. The ground tilt still limits the platform motion between $0.2-0.4$ Hz due to the lack of a low-frequency, low-noise rotation sensor on the platform. 

\begin{figure}[!h]
\begin{center}
\includegraphics[width=0.65\textwidth]{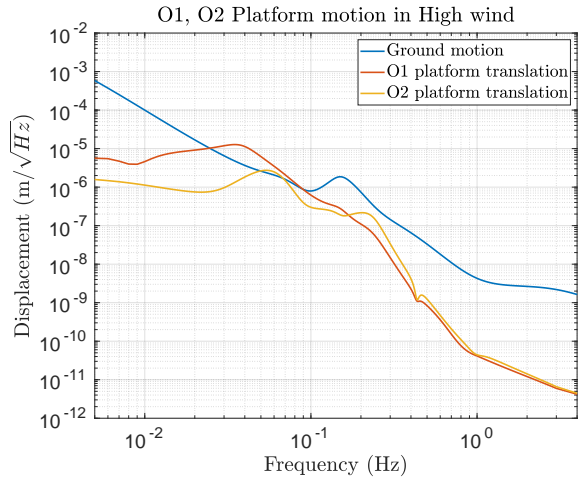}
\end{center}
\caption{Comparison of platform translation during high wind for O1 and O2.\label{PlatformCompare}}
\end{figure}
\subsection{O1 vs. O2}
Fig. \ref{PlatformCompare} shows a comparison of the platform motion during high winds for the O1 and O2 configurations. Above 450 mHz, the performance is almost identical. At 200 mHz the residual platform motion is increased by a factor $\sim2$ between the O2 and O1 configurations. However, the platform motion at $20-30$ mHz with O2 configuration is more than an order of magnitude less than that achieved by the O1 configuration. The increased performance is due to decreased tilt contribution and decreases the overall platform motion. This then decreases the motion that the downstream control loops must fight and thus decreases the chance of a lockloss.

\section{Comparison with Measured Data}
The model in the previous section allows a calculation of the platform motion for a given ground translation and tilt. In reality, we measure ground motion with seismometers, whose signal is a combination of translation and tilt, and ground rotation with the rotation sensors. Additionally, these sensors do not perfectly capture the input ground motion to the isolation platform.

It has been found that while most translations at these frequencies are coherent over large distances, tilts are coherent only over a few meters. This lack of coherence is thought to be due to non-linear deformation of the observatory floor. Thus, while tilt can be efficiently subtracted from a nearby seismometer, ground tilt measurements are not equivalent to the tilts experienced by the platform. The platforms and rotation sensors are separated by 5-7 meters and physical obstacles did not allow the sensors to be placed closer to the chambers. Additionally, the platform tilt is a combination of those experienced by its four widely spaced legs. These facts make precise prediction of the platform tilt performance for a given ground tilt measurement intractable and also removes the ability to use the ground rotation sensors for rotational sensor correction.

\begin{figure}[!h]
\begin{center}
\includegraphics[width=0.65\textwidth]{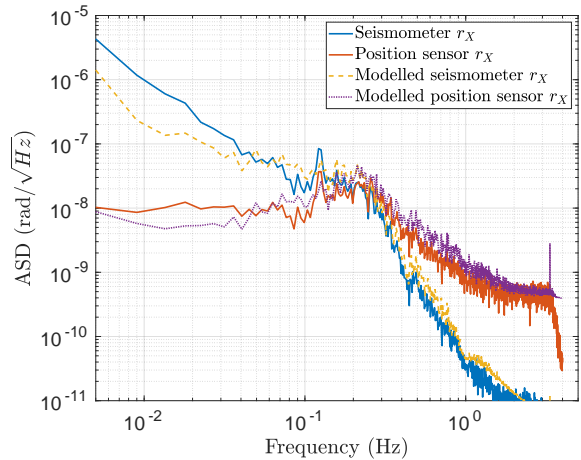}
\end{center}
\caption{Comparison of measured platform tilt to modelled tilt in high wind in the O2 configuration.\label{TiltDataWindyO2}}
\end{figure}

To test the model, we compare the predictions against measured platform motions during O2. Fig.~\ref{TiltDataWindyO2} shows the measured and modelled platform tilt during windy conditions. The first two curves are the measured T240 and CPS platform tilt sensors, which show the isolation above the blend frequency of $\sim$ $250$ mHz. Note that the CPS measures the platform motion relative to the ground while the T240 measures the inertial motion. The next two curves show the predictions of the model based on the measured ground tilt. The qualitative agreement is quite good, differing by less than a factor of $\sim$2-3 at most frequencies. The discrepancy is expected as there are differences between the measured ground tilt at the rotation sensor and the tilt experienced by the isolation platform.

\begin{figure}[!h]
\begin{center}
\includegraphics[width=0.65\textwidth]{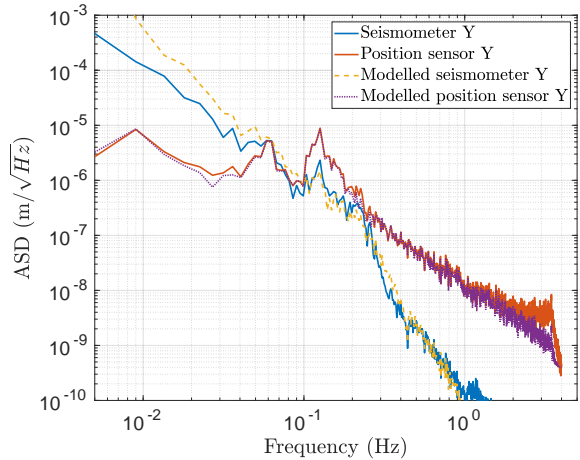}
\end{center}
\caption{Comparison of measured platform translation to modelled translation in high wind in the O2 configuration.\label{XDataWindyO2}}
\end{figure}

Similarly, Fig.~\ref{XDataWindyO2} shows the measured and modelled platform translation during windy conditions in O2. As before, the first two curves are the measured T240 and CPS translation sensors on the platform, which show the isolation above the blend frequency of $\sim$ $250$ mHz and the extra isolation due to the sensor correction at lower frequencies. The next two curves show the predictions of the model based on the measured ground translation and tilt. Once again, the qualitative agreement is quite good. 

\section{Filter tuning and performance measurements}

The figure of merit chosen for optimising the tilt-subtraction was the cumulative RMS velocity, integrated from high-frequencies to low-frequencies, of the tilt-corrected seismometer signal. There are a number of advantages when using the RMS velocity. First, it is an integrated measure of the spectrum that quantitatively shows whether a small improvement at some frequencies is worth some degradation at other frequencies. Second, the velocity determines key interferometer behaviours, including the coupling of scattered light. Third, there is no significant contribution to the RMS below approximately 1\,mHz, allowing a firm quantification of low-frequency noise injection for data stretches of a couple of hours. The tilt subtraction tuning was performed using a data-segment of 5-hours with moderate wind conditions.

\begin{figure}
\begin{center}
\includegraphics[width=0.65\textwidth]{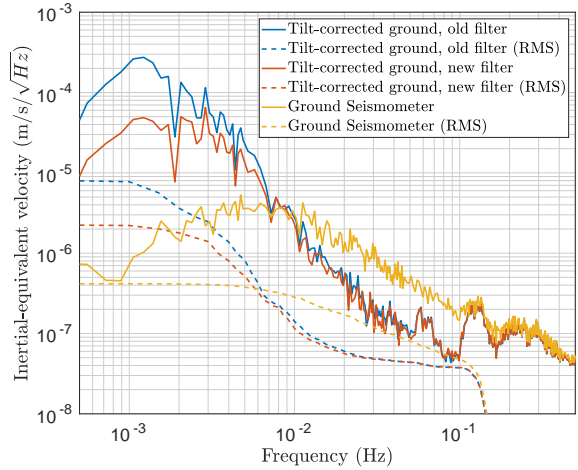}
\end{center}
\caption{The impact of optimising the tilt-subtraction filter. After turning, the tilt subtraction is very slightly improved, limited by the coherence of the tilt motion at the sensors, and at low frequencies the self-noise injection from the BRS is reduced by a factor of nearly 4 in RMS.}
\label{tiltFilterTuning}
\end{figure}

For the beam rotation sensor to be used most effectively, the tilt signal, corrected for its effect on translation, should simply be subtracted from the ground seismometer. However, there are several effects that can spoil the total performance: a calibration difference between the ground seismometer and the BRS, the self-noise of the BRS at low frequencies, the mechanical response of the BRS, and the natural AC-coupling of the seismometer. 

To compensate for each of these effects, we created a filter with separately tunable components: the overall gain, an AC-coupling frequency for the BRS at approximately 1\,mHz, the frequency and quality-factor of the plant-inversion, and the frequency and quality-factor of the seismometer-matching AC-coupling filter at approximately 8\,mHz. The effect of each filter component on the resulting RMS velocity was largely independent of the others, and they were optimised successively. 

Figure \ref{tiltFilterTuning} shows the improvement in performance after the tilt-subtraction filter was optimised. The dominant effect was matching the high-pass filter applied to the BRS with the AC-coupling filter in the seismometer. This both reduced the low-frequency noise injected by the BRS by a factor of approximately 4, and improved the phase-matching between the signals near 10\,mHz, slightly improving tilt-subtraction. The resulting filter was implemented and tested on new stretches of data, with similar effect. 

To quantify the downstream effect of tilt-subtraction on the motion of LIGO's isolated platforms, the CPS translation sensors are used as a measure of low-frequency motion. This approximation is valid for frequencies between 1\,mHz and approximately 0.1\,Hz. As previously, we look at spectral density and cumulative RMS of the velocity. Using this metric, the sensor-correction filter was tuned in a similar manner to the tilt-correction filter. There are many highly-coupled parameters in the sensor-correction filter, so it was not possible to grid-search for a global minimum, but the performance was slightly improved. 

\begin{figure}
\begin{center}
\includegraphics[width=0.65\textwidth]{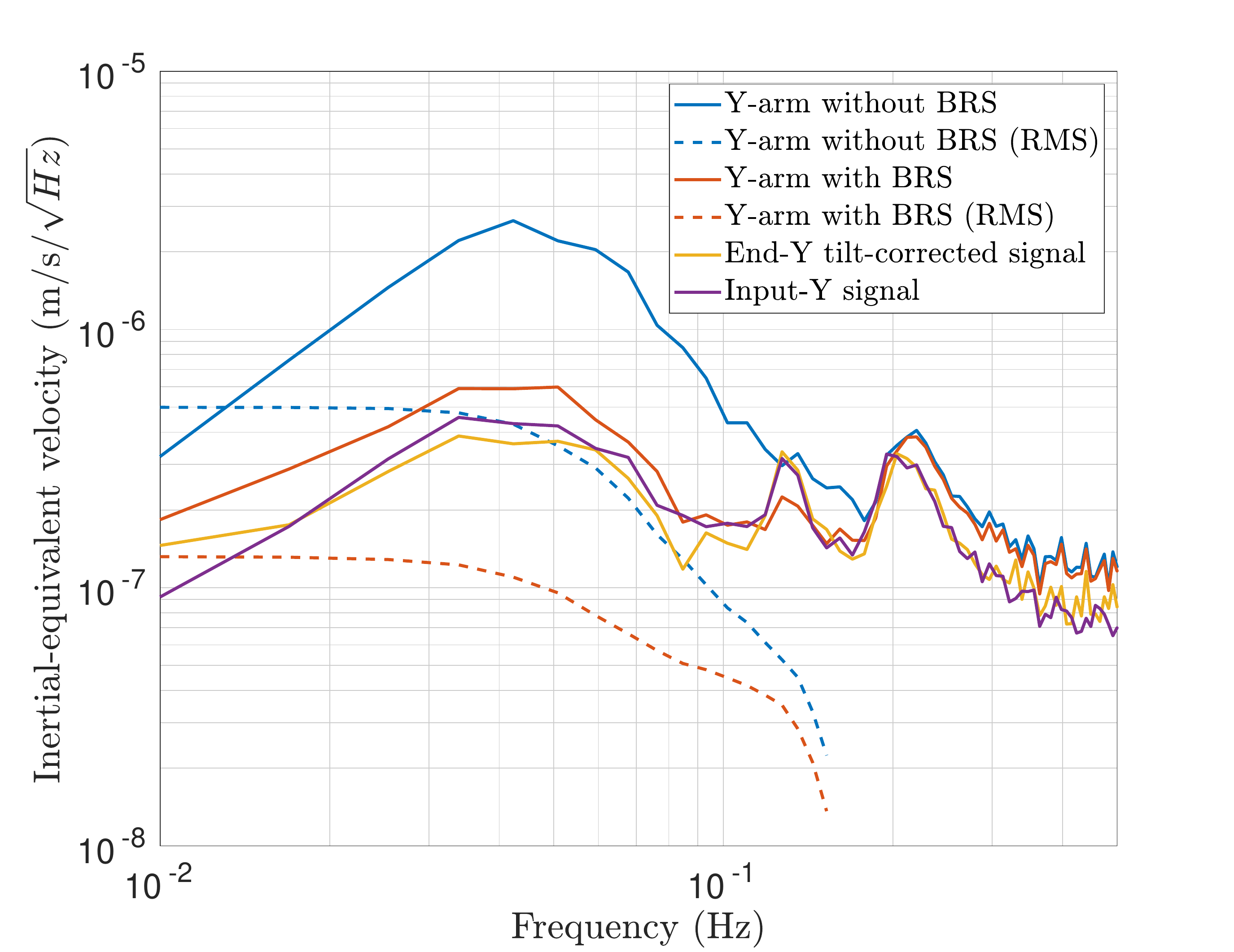}
\end{center}
\caption{The effect of BRS on the differential-velocity of the platforms in one of LIGO's arms during a period of high wind. The data was recorded over a 2-hour period when the wind speed averaged approximately 10\,m/s. The figure of merit is the Y-arm differential velocity when using the BRS (black dashed line) compared with the non-BRS case (dashed red line). The RMS is accumulated starting from 0.15\,Hz, indicated by the dashed vertical line.}
\label{yArmHighWind}
\end{figure}

\begin{figure}
\begin{center}
\includegraphics[width=0.65\textwidth]{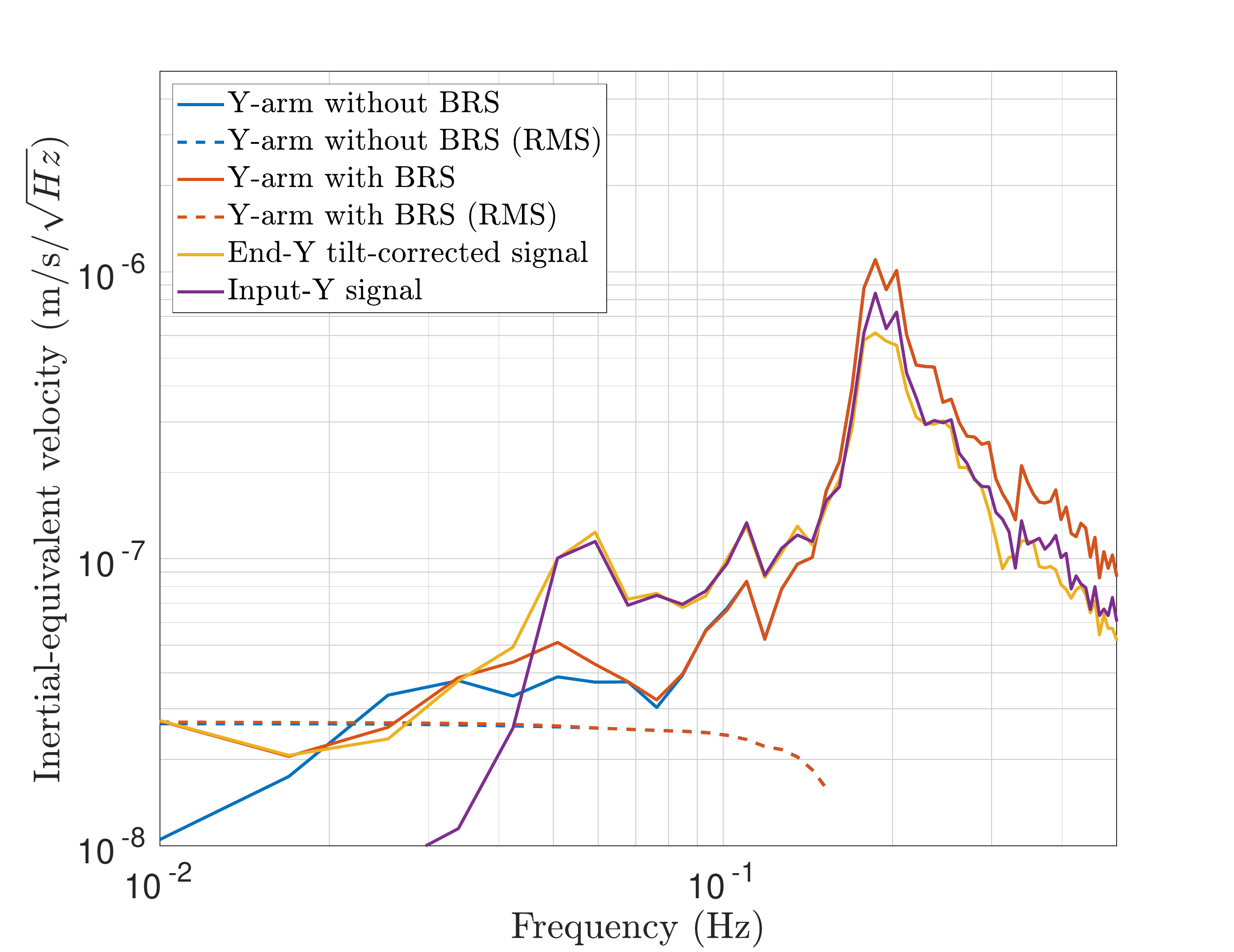}
\end{center}
\caption{The same sensors as in Fig.~\ref{yArmHighWind}, but for a 2-hour stretch with wind speeds of 1\,m/s. The two RMS curves are almost identical, proving that even in the lowest wind conditions, the BRS does not degrade performance or affect the common-mode rejection of translation.}
\label{yArmNoWind}
\end{figure}

To make the measurement more directly applicable to the complete interferometer, we also look at the difference between the 4\,km separated CPS signals, synthesising the spectral density of the arm-length fluctuations. Figures \ref{yArmHighWind} and \ref{yArmNoWind} show the effect of the tilt-subtraction on the rate-of-change of the length of the Y-arm for high and low wind conditions respectively. For the high-wind  case, the RMS velocity between the ISI platforms was reduced by a factor of 3.7. Interestingly, the differential velocity is lower than either of the individual velocities at the small peak near 0.13\,Hz, which is strong evidence for common-mode rejection of ground-translation after the tilt-subtraction process. For the low-wind case we see that the BRS adds no additional noise to the platform. The X-arm tilt-subtraction reduced the RMS velocity even further, by a factor of approximately 5, and the noise injection was similarly negligible during low winds.

\section{Impact on Duty-Cycle and noise}

As expected, the reduced platform motion under windy conditions has been anecdotally observed to make lock acquisition easier due to the decreased input motion to the control systems that keep the interferometer aligned. The impact of the O2 configuration on the duty cycle can be assessed by evaluating the fraction of the time the interferometer was locked during a given wind speed. 

\begin{figure}[!h]
\begin{center}
\includegraphics[width=0.65\textwidth]{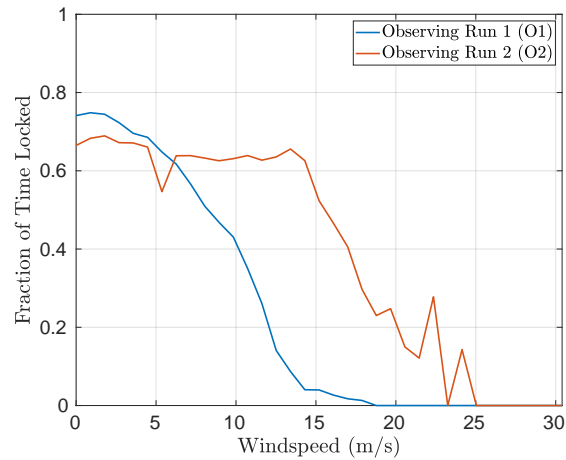}
\end{center}
\caption{Histogram of fraction of the time the interferometer was locked as a function of wind speed for the O1 and O2 configuration.\label{LockedComparison}}
\end{figure}

Histograms of the locked fraction are shown in Fig.~\ref{LockedComparison} for the O1 and O2 configurations.
The improvement to the duty cycle in O2 is evident for winds above 7 m/s. With the O1 configuration the locked fraction dropped monotonically with wind-speed, whereas in O2 it is nearly independent of wind-speed up to about 15 m/s, after which it falls off linearly due to residual tilt contamination. Decreases in duty cycle at low wind speed are due to differences in the observatory conditions between O1 and O2 that are independent of the seismic isolation scheme.

This increased robustness accounted for a decrease in the observing-time loss due to wind from 3.9\% during O1 to 0.3\% in O2. This equates to an additional 13.1 days of observing per year.

\pagebreak
\section{Summary}
The paper discusses the effect of ground tilt on the LIGO active seismic isolation. We have presented a simple analytical model to compare the impact of tilt on platform motion in two configurations and have shown the benefit of using a ground rotation sensor. This model shows qualitative agreement with measurements of the platform motion. The use of the ground-rotation-sensors has made the active isolation system less vulnerable to wind-induced tilt and improved the duty cycle of the LIGO Hanford Observatory significantly during wind speeds exceeding 7 m/s. Similar duty cycle improvements are expected to accompany the recent installation of ground rotation sensors at the LIGO Livingston Observatory. In the near future, the installation of on-platform compact-BRSs, which are currently in development, may decrease the tilt motion of the isolation platforms and further increase the performance of the low-frequency isolation.

\section{Acknowledgements}
The authors would like to thank Robert Schofield for his valuable comments. This work was carried out at the Laser Interferometer Gravitational-Wave Observatory (LIGO) Hanford Observatory (LHO) by members of LIGO laboratory and the LIGO Scientific Collaboration including University of Washington, Seattle, University of Birmingham, Standford University, and Massachusetts Institute of Technology. LIGO was constructed by the California Institute of Technology and Massachusetts Institute of Technology with funding from the National Science Foundation (NSF) and operates under Cooperative Agreement PHY‐0757058. Advanced LIGO was built under Award PHY‐0823459. Participation from the University of Washington, Seattle, was supported by funding from the NSF under Awards PHY-1607385, PHY-1607391, PHY-1912380, and PHY-1912514. Participation from Stanford University was supported by funding from the NSF under Award PHY-1708006. 
\section{References}
\bibliography{BRSRefs}{}

\appendix
\section{}
\begin{figure}[!h]
\begin{center}
\includegraphics[width=0.65\textwidth]{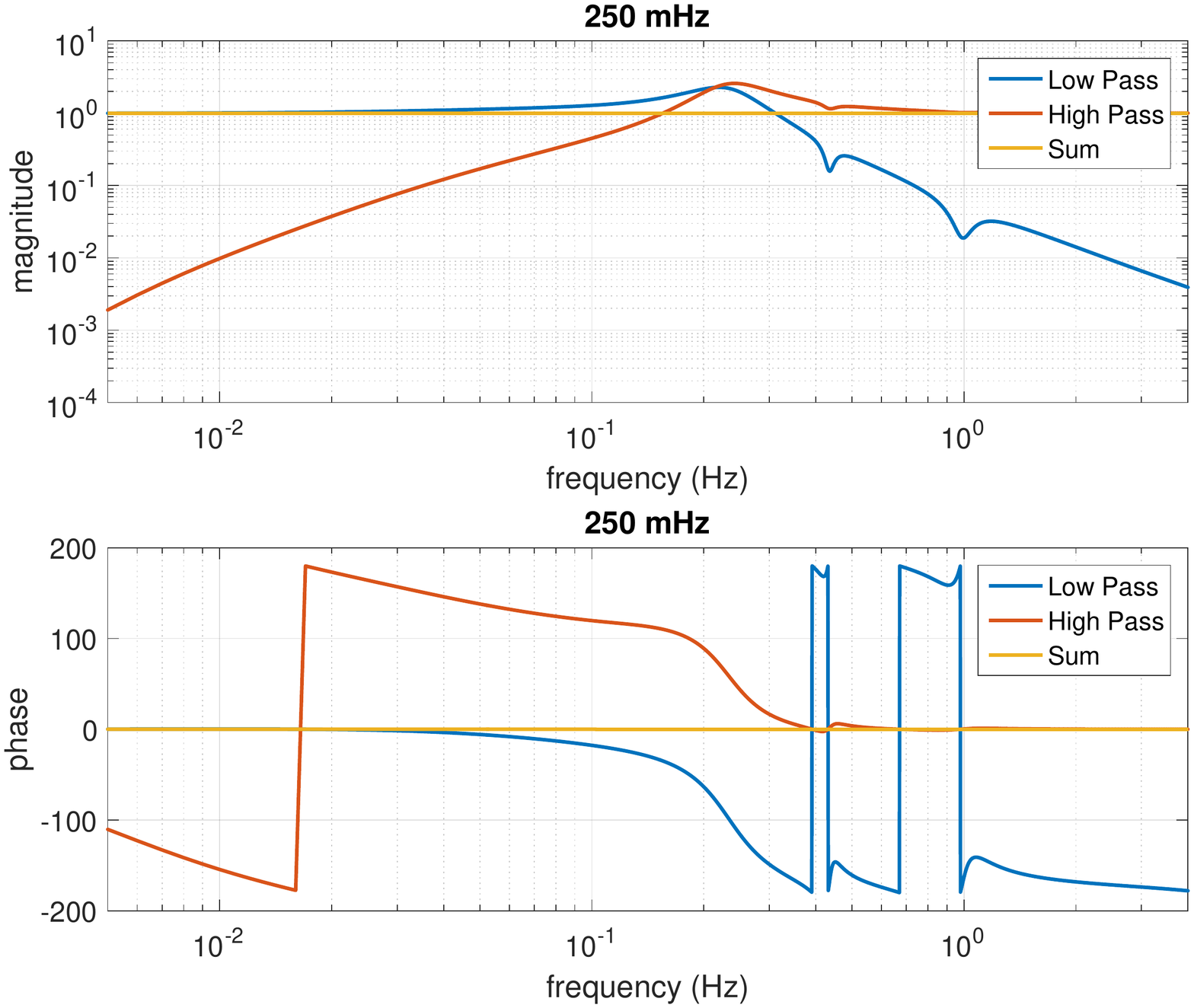}
\end{center}
\caption{250 mHz blend filters.\label{Q250}}
\end{figure}

\begin{figure}[!h]
\begin{center}
\includegraphics[width=0.65\textwidth]{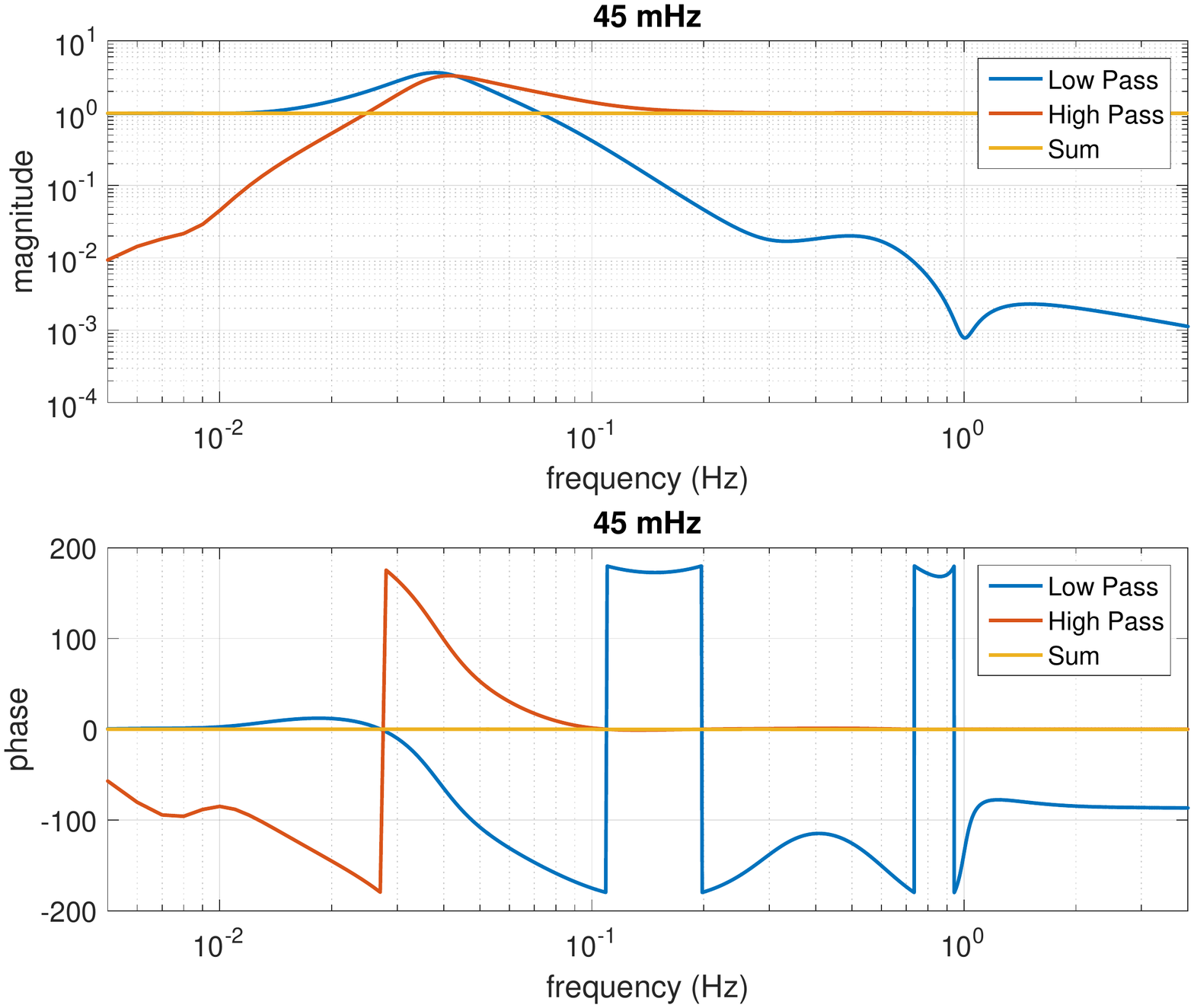}
\end{center}
\caption{45 mHz blend filters.\label{R45}}
\end{figure}

\begin{figure}[!h]
\begin{center}
\includegraphics[width=0.6\textwidth]{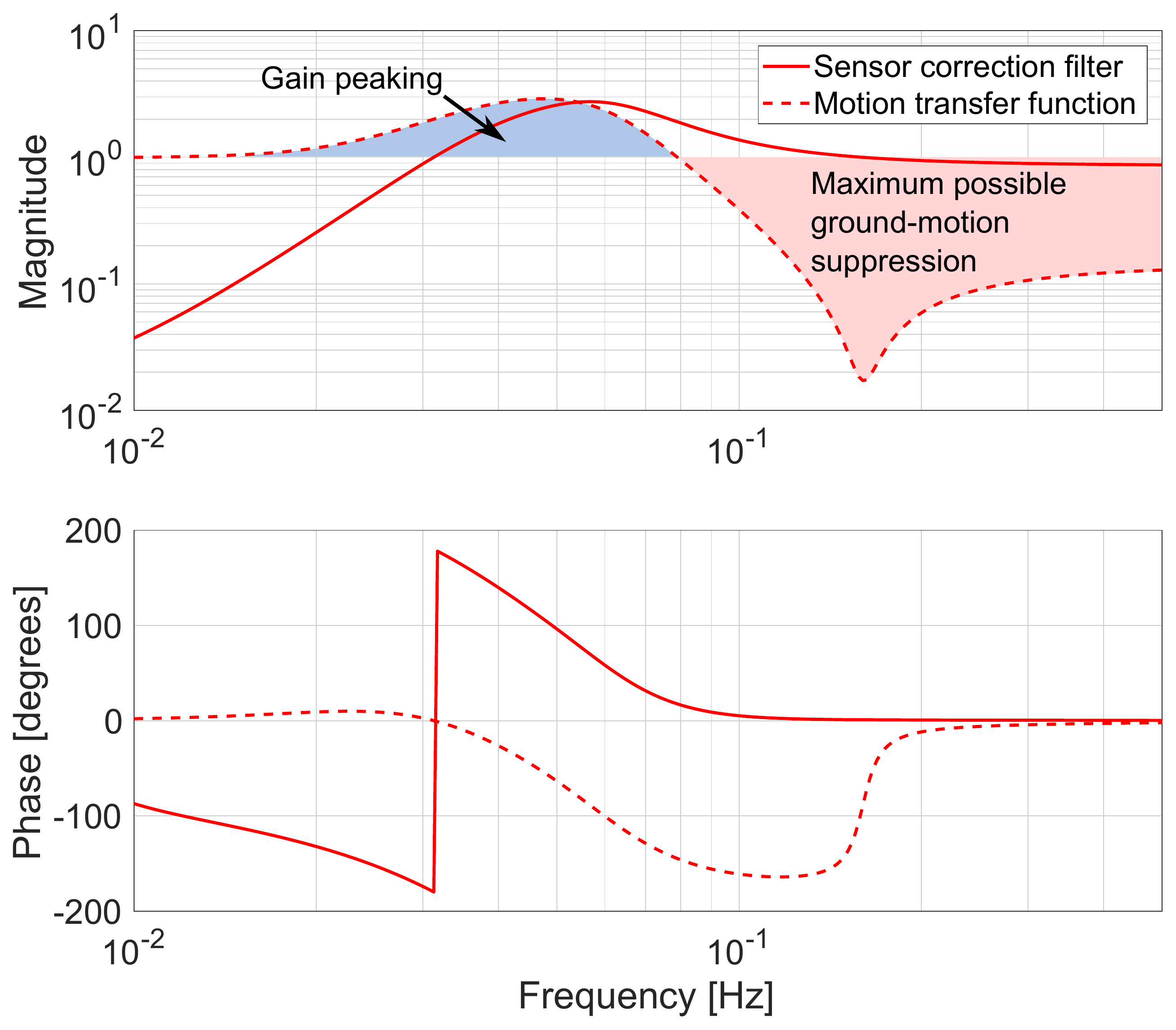}
\end{center}
\caption{Sensor correction filter showing the maximum possible ground-motion suppression.}
\label{SC}
\end{figure}

The two commonly used blend filters are shown in Fig.~\ref{Q250} and \ref{R45}, labelled as 250 mHz and 45 mHz blends. The sensor correction filter used in O2 is shown in Fig.~\ref{SC}.

\end{document}